\documentclass[conference]{IEEEtran}
\IEEEoverridecommandlockouts

\usepackage{cite}
\usepackage{amsmath,amssymb,amsfonts}
\usepackage{algorithmic}
\usepackage{graphicx}
\usepackage{textcomp}
\usepackage{mathtools}
\usepackage{multirow}
\usepackage{bm}
\usepackage{enumitem,kantlipsum}
\def\BibTeX{{\rm B\kern-.05em{\sc i\kern-.025em b}\kern-.08em
    T\kern-.1667em\lower.7ex\hbox{E}\kern-.125emX}}
\begin{document}

\title{Transformer-based Deep Learning Model for Joint Routing and Scheduling with Varying Electric Vehicle Numbers
\\
{\footnotesize}
\thanks{Hao Wang is supported in part by the Australian Research Council Discovery Early Career Researcher Award under Grant DE230100046.}}

\author{
    \IEEEauthorblockN{Jun Kang Yap\IEEEauthorrefmark{1}, Vishnu Monn Baskaran\IEEEauthorrefmark{1}, Wen-Shan Tan\IEEEauthorrefmark{2}, Ze Yang Ding\IEEEauthorrefmark{3}, Hao Wang\IEEEauthorrefmark{4}, David L. Dowe\IEEEauthorrefmark{4}}
    \IEEEauthorblockA{\IEEEauthorrefmark{1}School of Information Technology \IEEEauthorrefmark{2}\IEEEauthorrefmark{3}School of Engineering \IEEEauthorrefmark{2}Centre for Net-Zero Technology
    \\ \IEEEauthorrefmark{1}\IEEEauthorrefmark{2}\IEEEauthorrefmark{3}Monash University Malaysia, Bandar Sunway, Malaysia
    \\ \IEEEauthorrefmark{1}\IEEEauthorrefmark{2}\IEEEauthorrefmark{3}\{jun.yap, vishnu.monn, tan.wenshan, ding.zeyang\}@monash.edu}
    \IEEEauthorblockA{\IEEEauthorrefmark{4}Department of Data Science and AI, Faculty of IT
    \\ \IEEEauthorrefmark{4}Monash University, Clayton, Australia
    \\ \IEEEauthorrefmark{4}\{hao.wang2, david.dowe\}@monash.edu}
}

\maketitle

\begin{abstract}
The growing integration of renewable energy sources in modern power systems has introduced significant operational challenges due to their intermittent and uncertain outputs. In recent years, mobile energy storage systems (ESSs) have emerged as a popular flexible resource for mitigating these challenges. Compared to stationary ESSs, mobile ESSs offer additional spatial flexibility, enabling cost-effective energy delivery through the transportation network. However, the widespread deployment of mobile ESSs is often hindered by the high investment cost, which has motivated researchers to investigate utilising more readily available alternatives, such as electric vehicles (EVs) as mobile energy storage units instead. Hence, we explore this opportunity with a MIP-based day-ahead electric vehicle joint routing and scheduling problem in this work. However, solving the problem in a practical setting can often be computationally intractable since the existence of binary variables makes it combinatorial challenging. Therefore, we proposed to simplify the problem's solution process for a MIP solver by pruning the solution search space with a transformer-based deep learning (DL) model. This is done by training the model to rapidly predict the optimal binary solutions. In addition, unlike many existing DL approaches that assume fixed problem structures, the proposed model is designed to accommodate problems with EV fleets of any sizes. This flexibility is essential since frequent re-training can introduce significant computational overhead. We evaluated the approach with simulations on the IEEE 33-bus system coupled with the Nguyen-Dupuis transportation network. 
\end{abstract}


\section{Introduction} \label{intro}
As the world shifts its attention towards the adoption of renewable energy sources (RESs),
fast-acting and flexible resources are essential to maintaining load and generation balance in the grid due to the intermittent and uncertain RES outputs. In this context, battery energy storage systems (ESSs) are commonly used to shift surplus RES generation from periods of low demand to periods of high demand. More recently, researchers have explored the use of mobile ESSs instead due to their potential in providing greater flexibility than stationary ESSs by delivering energy across the transportation network (TN). However, their high upfront investment cost has prompted researchers to consider more readily available alternatives \cite{li2021routing}. One promising option is the electric vehicles (EVs), whose numbers have grown significantly in recent years and are expected to reach 145 million by 2030 \cite{InternationalEnergyAgency}. While primarily used for transportation, EVs typically remain parked and idle for many hours throughout each day, which presents itself as an opportunity to be leveraged as mobile ESSs when appropriate charging coordination is performed \cite{10.1145/3678717.3691247}. 

The EV charging coordination problem has been widely studied, with a wide variety of formulations being proposed. For instance, earlier research primarily focuses on coordinating EV charging and discharging to minimise grid operational costs while exclusively considering grid states \cite{yang2015unit, yang2019binary, rafique2021ev, rafique2022two, 9540895}. However, unlike traditional ESSs, EVs operate in both the grid and TN. This makes it inappropriate for EV charging coordination problems to focus exclusively on grid conditions since poor coordination on either network can potentially bring negative effects to the other. Therefore, recent research began expanding these formulations by incorporating TNs and optimising EV routing alongside with the grid operations \cite{TRIVINOCABRERA2019113, yao2021joint, zhao2022congestion, liu2022collaborative, aghajan2023optimal}.  

\begin{figure}[t]
\centerline{\includegraphics[width=\linewidth]{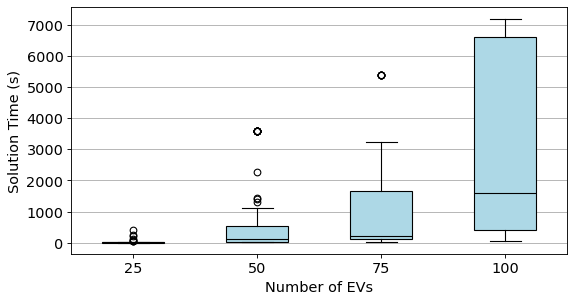}}
\caption{Standard box plots of solution times for 80 random samples of the formulated charging coordination problem with increasing number of EVs. Both the median and standard deviation of the solution times increase exponentially as the problem size increases.}
\label{graph}
\end{figure}

Similar to many power system related optimisation problems, mixed-integer programming (MIP) is commonly used to formulate the EV charging coordination problem. However, the combinatorial nature of MIP problems makes solving them inefficient, thus severely limiting their tractability in many real world tasks, especially time sensitive tasks such as the clearance of day-ahead and real-time energy markets \cite{bansal2022equilibrium}. As shown in Fig. \ref{graph}, solving MIP-based charging coordination problems using commercial solvers like Gurobi \cite{gurobi} can become prohibitively expensive as problem size grows. Recently, machine learning (ML), including deep learning (DL) and reinforcement learning (RL), has presented itself as a compelling solver alternative to solve MIP problems due to their near instantaneous inference time. Nonetheless, ML approaches often face challenges in balancing solution speed, feasibility and quality since they typically approximate solutions without taking problem constraints into account. Moreover, they also struggle to generalise to varying input and output dimensions resulting from problem structural changes. 

Therefore, this paper explores the charging coordination problem with a MIP-based, day-ahead electric vehicle joint routing and scheduling (EVJRS) stochastic optimisation problem. The problem seeks to optimise the operational cost of a distribution network (DN) equipped with distributed generators (DGs), solar photovoltaics (PVs) and EV charging stations (CSs). This is achieved by jointly optimising the EV routing decisions alongside with the DG and EV charging-discharging schedules. Solving these components in a single combined model can potentially yield better overall optimisation performance due to the availability of global information \cite{TRIVINOCABRERA2019113}. To overcome the computational challenges in solving the EVJRS problem, we propose to simplify the solution process by training a transformer-based DL model to predict the optimal binary solutions. This approach significantly prunes the solution search space, thus making the problem more tractable for a conventional solver. Additionally, we consider a practical scenario where the number of EVs can vary across different problem instances and design the transformer model to be able to dynamically generate the required number of binary variables for problems with EV fleet of any sizes, effectively reducing model re-training frequency. Therefore, the main contributions of this work are listed as follows:

\begin{enumerate}
\item \textbf{Fleet-Size-Agnostic Transformer-based Model for Joint Routing and Scheduling Problem:} 
We propose a transformer-based DL model to accelerate the solution process of a EVJRS problem, where the model is trained to predict the optimal binary solutions. We then utilise these predictions to prune the solution search space, thereby simplifying the problem and making it more tractable for a conventional MIP solver. In addition, the model is designed to handle problems with \textit{EV fleet of any sizes}, regardless whether that number of EVs has been seen during training. This adaptability significantly reduces the re-training frequency required, therefore lowering the overall training computational overhead. 
\item \textbf{Evaluation on Coupled Distribution and Transportation Network:}
We then evaluated the proposed transformer model's performance in assisting Gurobi to solve the EVJRS problem with a simulation on the IEEE 33-bus DN coupled with the Nguyen-Dupuis TN \cite{nguyen1984efficient}. The model is trained and tested on problem instances with fleet sizes within the range of 20 to 100 EVs. When compared to Gurobi, our approach reduced runtime by 98.1\% on average, while retaining 100\% feasibility and losing less than 0.0007\% of solution quality on the test dataset, all without having to re-train the model.

\end{enumerate} 

The paper is structured as follows. Section \ref{related} reviews the related works. Section \ref{problem} discusses the EVJRS problem and the proposed framework. Section \ref{method} details the proposed transformer model. Simulation results are discussed in Section \ref{case}. Lastly, the paper is concluded in Section \ref{conclude}.

\section{Related Works} \label{related}

As mentioned in Section \ref{intro}, there is a growing interest in applying ML algorithms to solve various MIP-based power system optimisation problems and researchers have proposed two main approaches: the direct and the hybrid assisting approach. In the direct approach, a more intuitive and natural approach is employed, where typically an end-to-end model is trained to directly output the final solution, thereby bypassing the computationally slow and expensive solver. This has been demonstrated in \cite{ojand2021q}, where the authors applied Q-learning to coordinate the thermal appliances together with the ESSs and solar PVs in an aggregated home microgrid to minimise the total electricity purchased from the main grid. Similarly, \cite{haider2024siphyr} trained a feed-forward network (FFN) to directly output the solution of a grid reconfiguration problem. However, this approach often becomes less effective as the problem complexity increases. This is because conventional ML algorithms typically struggle to enforce problem constraints, and the growing complexity of the prediction task can make ML predictions more prone to infeasibility. 

Consequently, many researchers favour the hybrid assisting approach, where typically a ML algorithm is used to either simplify a problem or to guide a solver, making the overall solution process more efficient. This hybrid approach has been demonstrated in \cite{yoldas2020optimal, gao2023hybrid, velloso2021combining, shekeew2023machine, o2023reinforcement, shekeew2023learning, wei2025data}, where the authors in \cite{yoldas2020optimal} applied Q-learning to first predict the actions of DGs and ESSs in a university microgrid before fine-tuning them with a solver. Besides that, a FFN was employed in \cite{gao2023hybrid} to determine the actions of uninterruptible load for a home energy management system while the remaining load actions were solved with a solver. Similarly, \cite{velloso2021combining} trained a FFN model to generate initial solution for the column and constraint generation algorithm to solve a security constrained optimal power flow problem. Last but not least, \cite{shekeew2023machine, o2023reinforcement, shekeew2023learning, wei2025data} had explored various approaches to predict the binary variables in a unit commitment problem.

While the hybrid approach does boost the solution feasibility, it is usually only achieved through heavy post-processing on the ML predictions. This is because a feasible simplification decision is still required for the solver to ensure final solution feasibility, and this serves as a key motivation for this work. Additionally, in practice, many real world problems may exhibit dynamic structures. In the context of EV charging coordination problems, the EV fleet size may change, leading to a change in number of decision variables. This presents a significant challenge to applying ML to solve power system optimisation problems since even minor modifications to problem structures may necessitate re-training, leading to a considerable amount of training overhead. These shortcomings will be addressed in this work.

\section{EVJRS Formulation} \label{problem}
\subsection{Problem Description}

\begin{figure*}[t]
\centerline{\includegraphics[width=\textwidth]{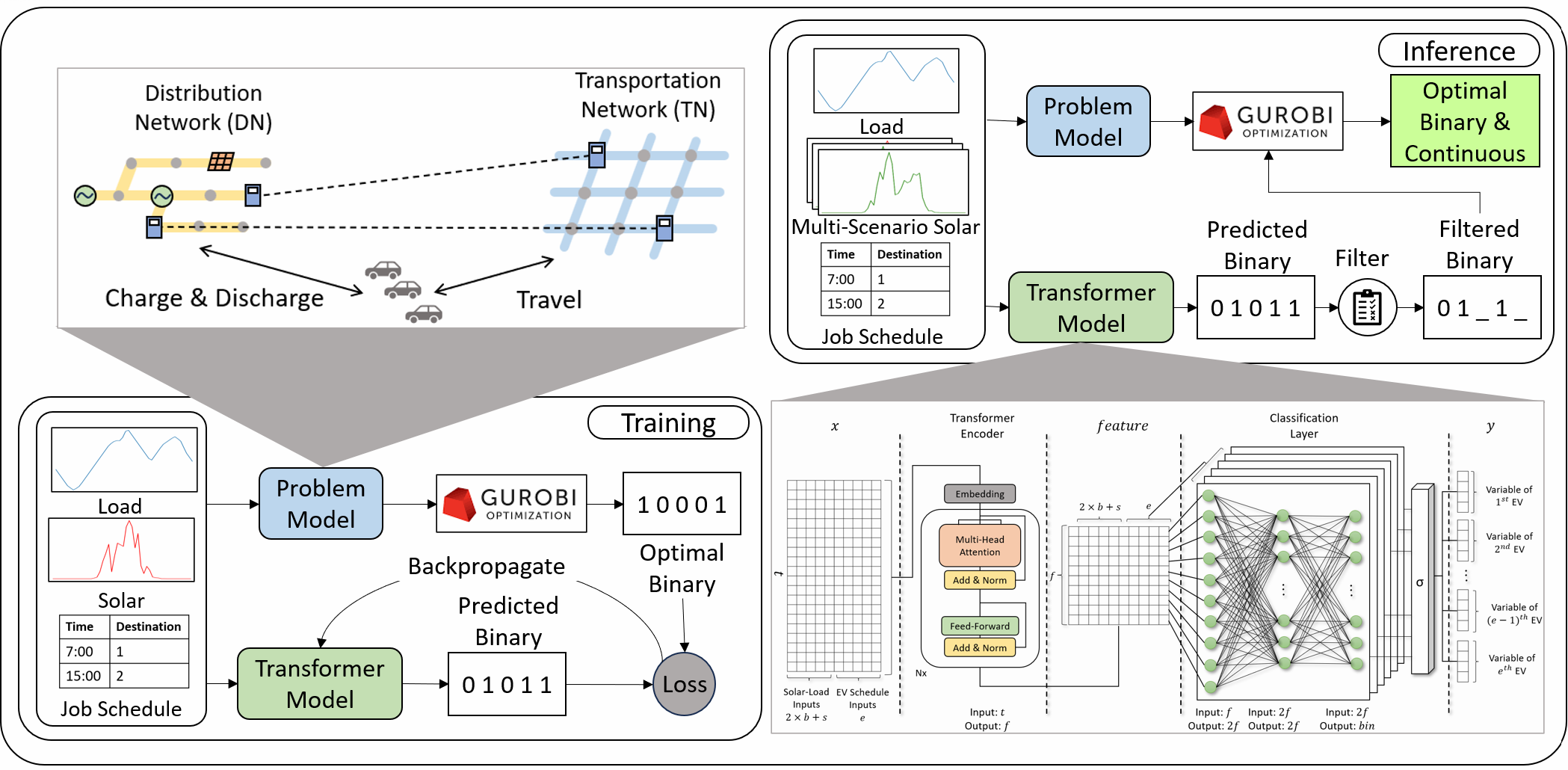}}
\caption{Training and inference framework of the proposed DL solution methodology. An overview of the problem model and the proposed transformer-based DL model are also provided here.}
\label{framework}
\end{figure*}

In this section, we first introduce the day-ahead EVJRS stochastic optimisation problem explored in this work. A simplified illustration of the problem is provided in Fig. \ref{framework}. The problem first considers a DN operator offering contracts to the EV owners, allowing them to participate in a program designed to support the operations of a DN equipped with DGs and solar PVs. In return of the services provided, the DN operator compensates the EV owners by providing some discharging earnings. Before the start of every day, the EV owners will first be required to submit their job schedules to the DN operator. The operator then uses a set of forecasted solar generation, load demand and the submitted job schedules to find a solution that minimises the DN operational cost by scheduling the DG operations and EV charging-discharging while optimally routing the EVs to fulfil their job schedules. The objective of the problem is to minimise the total generation cost and net EV charging cost as shown in \eqref{eq1},
\begin{align}
    \begin{split} 
    \min\; 
    \sum_{sc\in SC} \sum_{u\in G} \sum_{t\in T} c^f (P^{g}_{sc,u,t})p_{sc} +
    \\ \sum_{sc\in SC} \sum_{k\in K} \sum_{i\in CS} \sum_{t\in T} [c^c(P^{c}_{sc,k,i,t}) - c^d(P^{d}_{sc,k,i,t})]p_{sc}. \label{eq1}
    \end{split}
\end{align}
\noindent where $SC$, $G$, $T$, $K$ and $CS$ represents the set of scenarios, DGs, timesteps, EVs and CSs respectively. Moreover, $c^f$, $c^c$ and $c^d$ are the generation fuel cost, EV charging cost and discharging earnings. Lastly, $P^g$ is the generation power, $P^c$ and $P^d$ are the EV charging and discharging power.

\begin{figure}[t]
\centerline{\includegraphics[width=0.85\linewidth]{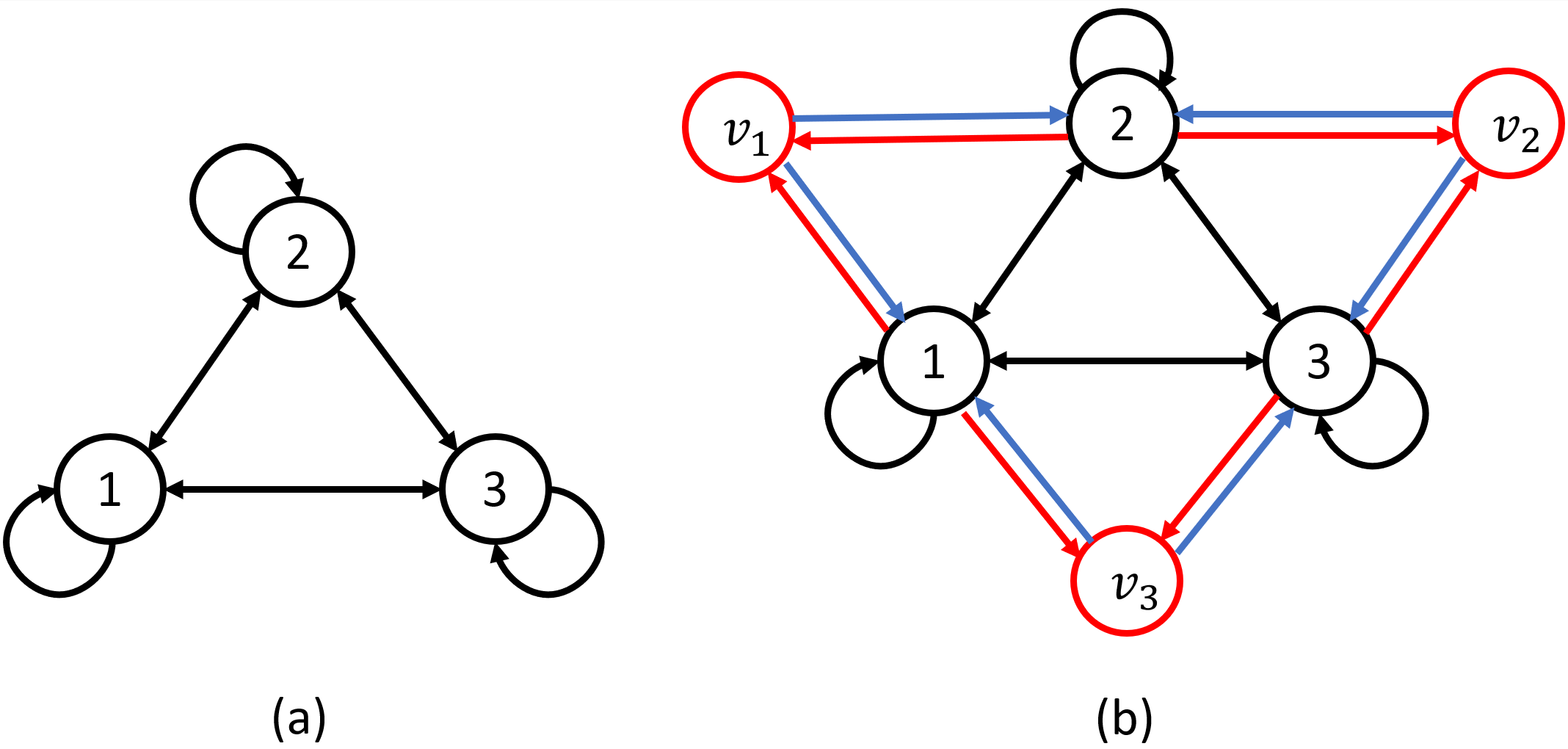}}
\caption{The original TSN (left) proposed in \cite{7024941} and the modified TSN (right).}
\label{tsn}
\end{figure}

The constraints considered in our EVJRS problem are categorised into three main categories: EV routing constraints, EV power constraints and the DN operational constraints. A brief discussion of the constraints will be provided here but more detailed formulation can be found in our previous work \cite{yap2025joint}. The constraints are listed as follows:

\textbf{EV Routing Constraints}: The EV routing constraints mainly enforce
the EV routing rules in the TN while also ensuring each EV fulfils its job schedules. These constraints are formulated using a modified time-space network (TSN) that takes into account changing trip times under different traffic conditions. The modified TSN model is illustrated in Fig. \ref{tsn}, where the congestion trip times are modelled through the addition of a virtual congestion node (red nodes). By enabling the congestion arcs (red arcs) and disabling the normal arcs (black arcs) during a congestion, an alternate path with longer trip time can be activated. Conversely, the opposite can be done when there is no congestion to re-enable the original route between the original nodes (black nodes). However, the arcs exiting the virtual congestion nodes (blue arcs) must always remain enabled to ensure the EVs are not trapped in the congestion state. 

The EV routing constraints are formulated in \eqref{eq2}-\eqref{eq5}, where the modified TSN is specifically modelled through constraints \eqref{eq2}-\eqref{eq4}. The behaviour of the normal and congestion arcs under different traffic conditions is modelled with \eqref{eq2}, where $s \in S$ is the set of timespans between each timestep, $CA$ and $NCA$ are the sets of arcs in the TN that are enabled when there is a traffic congestion and no congestion respectively, $I$ is the EV arc selection binary variable, and $j$ is the binary that indicates whether there is a traffic congestion occurring. Constraint \eqref{eq3} then ensures that each EV will only have a single travelling behaviour during each timespan. Next, constraint  \eqref{eq4} enforces EV flow conservation, where $[A^{from}, A^{to}]$ is the set of pairs of arc sets with the same origin and destination node respectively. Lastly, constraint \eqref{eq5} ensures that each EV adheres to the assigned job schedules. Each schedule $\xi \in \Xi$ is represented as a tuple $(k_{\xi}, i, s_{\xi})$, specifying that EV $k_{\xi}$ must arrive at node $i$ during timespan $s_{\xi}$, where $A^{i+}_{\xi}$ is the set of arcs that have node $i$ as the origin node. 
\begin{gather}
        \begin{dcases*} \label{eq2}
            \sum_{ij\in CA} I_{sc,k,ij,s} = 1, & \text{if } $j_s = 1$ \\
            \sum_{ij\in NCA} I_{sc,k,ij,s} = 1, & \text{otherwise}
        \end{dcases*}  ,\forall{sc}, \forall{k}, \forall{s},
\\
        \sum_{ij\in CA\,\cap\,NCA} I_{sc,k,ij,s} = 1 ,\forall{sc}, \forall{k}, \forall{s}, \label{eq3}
    \\
    \begin{split}
    \sum_{ij\in A^{from}} I_{sc,k,ij,s+1} = \sum_{ij\in A^{to}} I_{sc,k,ij,s} , \\ \forall{sc}, \forall{k}, \forall [A^{from}, A^{to}], \forall{s=1,2,...|S|-1}, \label{eq4}
    \end{split}
\\
    \sum_{ij \in A^{i+}_{\xi}} I_{sc,k_{\xi},ij,s_{\xi}} = 1, \forall{sc}, \forall{\xi \in \Xi = (k_{\xi}, i, s_{\xi})}, \label{eq5}
\end{gather}
\textbf{EV Power Constraints}:
Next, the EV power constraints mainly model the EV charging and discharging rules. This includes the charging and discharging constraints at CSs as well as the energy losses while travelling. The EV power constraints also ensure the EV energy balance. These constraints are defined in \eqref{eq6}-\eqref{eq11}, where \eqref{eq6} ensures that charging or discharging can only be performed when an EV is idling at a CS, where $I^{c}$ and $I^{d}$ are the binary variables that indicate the EVs' charging and discharging statuses respectively, and the index $ii$ refers to the arcs where the EVs will be idling at CS $i$. Constraint \eqref{eq7} ensures that power consumption will only occur if the EV is travelling, where $P^{m}$ denotes the EV power consumption while travelling and $P^{move}$ is the unit power consumption of each arc. Next, \eqref{eq8}-\eqref{eq9} limit the maximum and minimum EV charging and discharging rates respectively. Similarly, constraint \eqref{eq10} enforces the EVs' upper ($E^{max}$) and lower ($E^{min}$) energy bounds. Finally, the energy balance of each EV is maintained by \eqref{eq11}, where $\eta$ indicates the inefficiency during charging and discharging.
\begin{gather}
    \begin{split}
    I^{c}_{sc,k,i,s+1} + I^{d}_{sc,k,i,s+1} \leq I_{sc,k,ii,s}, \\ \forall{sc}, \forall{k}, \forall{i\in CS}, \forall{s}, \label{eq6}
    \end{split}
\end{gather}
\begin{gather}
    P^{m}_{sc,k,s+1} = \sum_{ij\in NSA} P^{move}(I_{sc,k,ij,s}), \forall{sc},\forall{k}, \forall{s},  \label{eq7}
\\
    \begin{split}
        0 \leq P^{c}_{sc, k, i, t} \leq P^{ev,max}_{k}(I^{c}_{sc,k,i,t}),   \forall{sc}, \forall{k}, \forall{i\in CS}, \forall{t},
    \end{split} \label{eq8} 
\\
    \begin{split}
        0 \leq P^{d}_{sc, k, i, t} \leq P^{ev,max}_{k}(I^{d}_{sc,k,i,t}),  \forall{sc}, \forall{k}, \forall{i\in CS}, \forall{t},
    \end{split} \label{eq9} 
    \\  
        E^{min}_{k} \leq E_{sc,k,t} \leq E^{max}_{k}, \forall{sc}, \forall{k}, \forall{t}, \label{eq10}
\\
    \begin{split}
        E_{sc,k,t} = E_{sc,k,t-1} + \frac{24}{|T|}[(1-\eta)\sum_{i\in CS}P^{c}_{sc, k, i, t} - \\ (1+\eta)(\sum_{i\in CS}P^{d}_{sc, k, i, t} + P^{m}_{sc,k,t})] ,\forall{sc}, \forall{k}, \forall{t=2,...,t},
    \end{split} \label{eq11} 
    \end{gather}
\textbf{DN Operational Constraints}:
Furthermore, the DN operational constraints model the DG generation as well as the line power and bus voltage constraints. The line power and bus voltage constraints are modelled using the LinDistFlow \cite{du2020interval} formulation. Last but not least, the uncertainty of solar generation is taken into account using scenario-based modelling.

\subsection{Proposed Framework}

The proposed DL solution methodology is briefly discussed in this section. Fig. \ref{framework} provides a summary of the training and inference framework of the said approach. We propose training the transformer-based DL model using supervised learning. In order to construct the labelled training dataset, we first solve a large set of the EVJRS problem instances with Gurobi to determine the optimal binary solutions. The model is then trained to predict these optimal binary solutions during inference. Since it is possible that the predictions can be infeasible, a thresholding filter is applied on the predictions to remove bits that are likely to be incorrectly predicted. The filtered predictions are subsequently fed into Gurobi to prune the solution search space, thus making the EVJRS problem more tractable for Gurobi.

\section{Fleet-Size-Agnostic Transformer model} \label{method}

\subsection{Training Dataset}

As mentioned in Section \ref{problem}, a labelled dataset is required to train the proposed transformer-based DL model. We employed similar procedures as our previous work \cite{yap2025joint} to construct the dataset and a summary of the said procedures are as follows. 

The EVJRS problem in this work is solved on a daily basis, with minor variation in parameters, namely the solar generation, load demand and EV job schedules. Therefore, we first construct a problem database by mix-and-matching random combination of solar generation, load demand and EV job schedules to build problem instances under different operating conditions. We then use Gurobi to solve these problems to obtain the optimal binary solutions (EV routing, charging and discharging decisions). Pre-processing is then carried out to obtain the DL inputs -- solar generation $(S \in \mathbb{R}^{\:sc \times s \times t})$, load demand $(PQ \in \mathbb{R}^{\:(2 \times b) \times t})$ and EV job schedules $(J \in \mathbb{R}^{\:e \times t})$, where $sc$, $s$, $t$, $b$ and $e$ indicate the number of scenarios, solar PVs, timesteps, DN buses and EVs respectively. The outputs would then be the corresponding optimal binary solutions. Note that we eased the labelling process by training the DL model using deterministic version of the EVJRS problem instead. The trained model is then used to make $sc$ number of predictions during inference before combining them to obtain the full binary solution.

\subsection{Feature Extraction -- Encoder}
Using the generated dataset, the prediction task can essentially be formulated as a multivariate time-series multi-label classification problem. Accordingly, we designed a transformer-based DL model for the task at hand. The choice for a transformer-based DL model is mainly motivated by its attention mechanism, which is able to capture long term dependencies in sequential data, as well as its inherent ability to process inputs with variable length. A key property of the EVJRS problem in Section \ref{problem} is that the binary decision variables are all associated with the EVs, essentially causing their numbers to scale proportionally with the EV fleet size. We leverage this property and designed the transformer model to adapt to problem instances with EV fleet of any sizes.

\begin{figure}[t]
\centerline{\includegraphics[width=\linewidth]{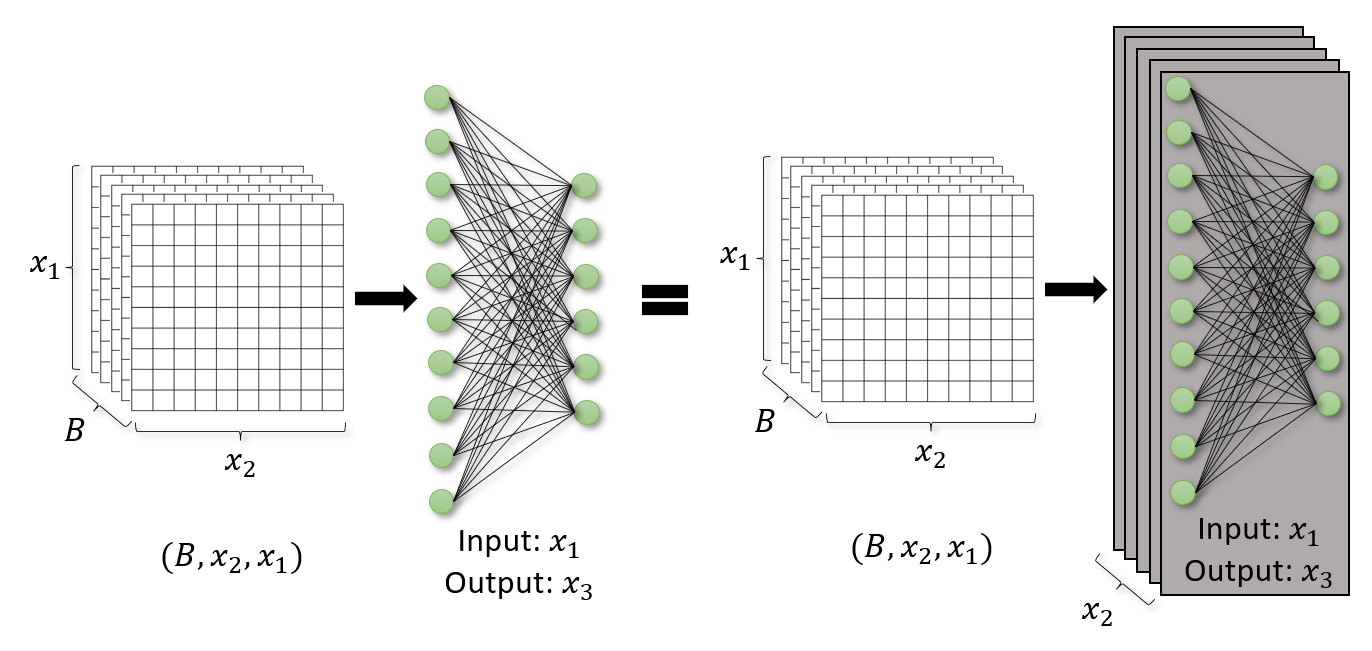}}
\caption{FFN mechanism for variable input dimensions.}
\label{mechanism}
\end{figure}

Like many DL model architectures, our transformer-based model also composes of two main components: the feature extraction and classification modules. We employed two layers of the standard transformer encoders \cite{vaswani2017attention} for feature extraction, followed by a FFN classification module. An overview of the proposed transformer model's architecture is provided in Fig.~\ref{framework}. In the feature extraction module, an embedding layer is first used to map the input to a more suitable representation for processing. The embedding layer is implemented using a standard FFN as shown in \eqref{eq12a}-\eqref{eq13},
\begin{gather}
    h_{0} = Concat(S,PQ,J), \label{eq12a}
    \\
    h_l = Embedding(h_{0}),  \label{eq12}
    \\
    Embedding(h_0) = ReLU(W_lh_0 + b_l) \label{eq13}
\end{gather}
where $h_l$ is the output from hidden layer $l$, and $W$ and $b$ are the trainable weights and biases of the FFN layer. Although FFN layers typically expects input tensors of shape $(B,x_1)$, where $B$ is the batch size, they can also handle inputs with additional dimensions. For example, given an input with a dimension of $(B,x_2,x_1)$, the FFN expecting inputs $(B,x_1)$ would be applied independently across the $x_2$ dimension, allowing the FFN layer to process variable-sized inputs. This mechanism has been illustrated in Fig. \ref{mechanism}.

The embeddings can then be passed through a series of transformer encoders with a multi-head self-attention mechanism (MHSA) for feature extraction as defined in \eqref{eq14}-\eqref{eq17},
\begin{gather}
    Attention(Q,K,V) = softmax(\frac{QK^T}{\sqrt{d_k}})V, \label{eq14}
    \\
    Q, K, V = h_lW^{Q,K,V}, W^{Q,K,V} \in \mathbb{R}^{\:|h_l| \times d_k}, \label{eq15}
    \\
    \begin{split}
    MHSA(h_l) = Concat(head_1, ..., head_{i})W^M, \\
    \text{where } head_i = Attention(Q,K,V),
    \label{eq16}
    \end{split} 
    \\
    h_{l+1} = LayerNorm(h_l + MHSA(h_l)), \label{eq17}
\end{gather}
where \eqref{eq14}-\eqref{eq15} represents the self-attention mechanism, \eqref{eq16} performs the multi-headed aggregation and \eqref{eq17} performs the ``Add and Norm" operation in the transformer encoder. The transformer encoder is inherently designed to process inputs with variable length. Typically, it expects an input tensor of shape $(B,T,F)$, where $T$ is the sequence length and $F$ is the feature dimension. Rather than using $T$ as the time dimension, we repurpose it for the feature ($2 b + s + e$) and $F$ dimension for the time ($t$), since we have varying feature length and fixed timesteps instead. Since the encoder only acts on the $F$ dimension, the $2 b + s + e$ can be preserved and passed to the classification module. However, a limitation of this approach is that the attention mechanism now captures dependencies across the feature dimension instead, which may reduce the model's ability to capture the temporal dependencies.

\subsection{Classification -- FFN}

The extracted features are then passed to a classification module to generate the binary predictions. Following the same mechanism of the feature embedding layers mentioned previously, a series of FFN layers with rectified linear unit (ReLU) activation function was chosen as the classification modules. We stacked three FFN layers as the classification module and passed the output to a sigmoid activation to produce the binary predictions a shown in \eqref{eq18}, where $\hat{y}$ is the binary predictions. It is important to note that a softmax activation should not be used since the problem is a multi-label prediction task and not a multi-class .
\begin{gather}
    \hat{y} = \sigma(h_l) = \frac{1}{1+e^{-h_l}}  
    \label{eq18}
\end{gather}

\subsection{Post-processing}

Although the DL model is only trained to assist Gurobi in accelerating the solution process of the EVJRS problem by predicting the binary variables, it is still essential that the predictions are feasible so that the solver can be guided to a valid solution space. To improve the final solution feasibility, we apply a threshold-based filter to extract predictions with higher confidence during inference. This approach leverages the fact that the sigmoid function outputs a probability that indicates how close a prediction is to the value of one. By retaining predictions closer to the extremes (i.e., probability near 0 or 1), we can extract binary bits that are more confidently predicted.
\begin{equation}\label{eq19}
        {p_t}^{1/0}=
        \begin{cases} 
            \hat{y} & \text{if $y$ = 1} \\
            1-\hat{y} & \text{otherwise,}
        \end{cases}
    \end{equation}
The thresholds used to filter both the classes (class 0 and 1) are computed by first taking the prediction probability (PP) of each binary label using the formulation in \eqref{eq19} \cite{ridnik2021asymmetric}. The mean PPs of both classes are then taken as the thresholds to select variables to be retained during inference. The selected variables are then used to prune the solution search space, making the EVJRS problem more tractable for Gurobi.

\subsection{Evaluation Metrics}

We used a total of five metrics to evaluate the performance of the trained transformer-based DL model. Our evaluation focuses on two aspects: the prediction performance of the model and also its ability to assist Gurobi in solving the EVJRS problem. The prediction performance is evaluated using class-specific accuracies ($ACC_{0}$ and $ACC_{1}$). This is mainly to mitigate bias since there is a significant imbalance between these two classes in a multi-label classification problem. However, we noticed that prediction accuracy alone is often insufficient to fully capture the model's ability to assist Gurobi. Therefore, we use mean percentage of computational time reduction ($\bar{r}$), mean percentage of loss of solution quality ($\bar{l}$) and percentage of feasible samples ($feas)$ as calculated with \eqref{eq20}-\eqref{eq22} to evaluate the model's solution performance, 
\begin{gather} 
    \bar{r} = \frac{1}{N} \left(~\displaystyle\sum\limits_{n=1}^{N} \left(1 - \frac{t^{a}_{n}}{t^{g}_{n}}\right)~\right) \times 100\%, \label{eq20}
    \\
    \bar{l} = \frac{1}{N} \left(~\displaystyle\sum\limits_{n=1}^{N} \left(\frac{v^{a}_{n} - v^{g}_{n}}{v^{g}_{n}}\right)~\right) \times 100\%, \label{eq21}
    \\
    feas = \frac{N_{DL}}{N} \times 100\%, \label{eq22} 
\end{gather}
where $N$ is the number of test samples, $t^{a/g}$ and $v^{a/g}$ indicates total solution time and objective value achieved by the DL assisted approach ($a$) and Gurobi ($g$) respectively. Lastly, $N_{DL}$ is the number of test samples where the DL assisted Gurobi managed to successfully return a feasible solution. Note that we assume all infeasible samples are solved using Gurobi to obtain a more accurate $\bar{r}$ and $\bar{l}$ measurement.

\section{Numerical simulation} \label{case}

\subsection{Simulation Setup}
\begin{figure}[t]
\centerline{\includegraphics[width=0.9\linewidth]{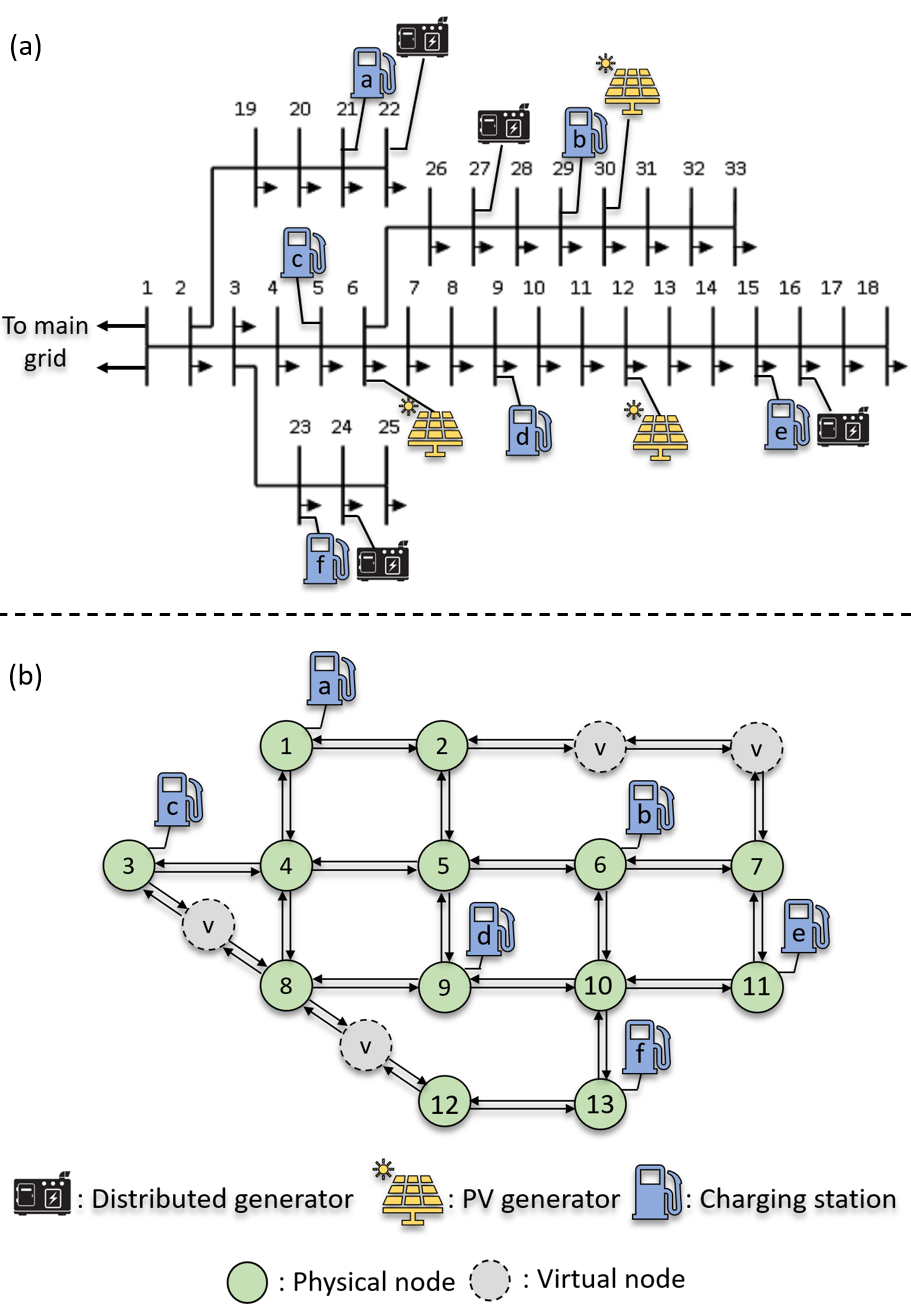}}
\caption{(a) IEEE 33-bus and (b) Nguyen-Dupuis transportation network.}
\label{network}
\end{figure}

We conducted a simulation study on the IEEE 33 bus system coupling with the Nguyen-Dupuis TN as shown in Fig.~\ref{network}, to evaluate the performance of the proposed transformer-based DL model. The simulation was developed entirely in Python, using Gurobi for the EVJRS model and PyTorch for training the transformer model. All computations were executed on a machine with a Intel i9-12900K CPU, NVIDIA RTX A5000 24 GB GPU and 128 GB of RAM. 

We fixed the number of solar generation scenarios to 5 and assigned equal occurrence probability to them. Furthermore, as outlined in Section \ref{problem}, we assume that each EV will have two job schedules throughout the day. Therefore, we first randomly assigned a start and ending position node $(o,d)$ chosen from the set $\{(1,11), (1,13), (3,11), (3,13)\}$ to each EV. Based on the assigned start and ending position, the two schedules will then be set according to the following rules:
\begin{itemize}
\item The first schedule with a random destination node selected from the node set $\{6,7,9,10,12\}$, at a random time between the period of 6:00 AM to 12:00 PM, 
\item followed by a second schedule at a random destination node from the set $\{2,4,5,8\}$, between 2:00 to 10:00 PM if the starting position was 1 and vice versa if it was 3.
\end{itemize}

\noindent The code to replicate this work is available at https://github.com/YapJunKang/TransformerEVJRS.

\subsection{Evaluation}
In addition to evaluating the computational advantages provided by the proposed approach, we also investigated how the transformer model performs when trained with datasets exhibiting different degrees of unseen scenarios. To simulate this, we constructed a total of four distinct training dataset, each comprising problem instances with different EV count multipliers. Each dataset was then used to train a separate transformer model. More specifically, we first restrict the number of EVs in the EVJRS problem to the range of 20 to 100. We then define a multiplier $m \in \{5,10,5,20\}$, which determines the step size of the EV counts of the problems in each dataset. For example, for the dataset with $m = 5$, it will include problem instances with EV counts of $e \in \{20, 25, ..., 100\}$ (i.e., multiplier of 5). 

Using the procedure mentioned in Section \ref{method}, we generated and labelled 800 problem instances for each of the training datasets, where the 800 samples are evenly distributed across the corresponding EV counts in the dataset. Each dataset is then normalised and split into training and validation subsets (90\% and 10\%). To ensure a fair evaluation on the trained transformer models, we then constructed a separate testing dataset consisting of problems with EV counts unseen by any of the models. We note that all the training datasets only consist of deterministic versions of the EVJRS problem, while the testing dataset consists of the full stochastic version. The details of each dataset are provided in Table \ref{dataset}.

\begin{table}[t]
\caption{Training dataset attributes with varying degree of unseen scenarios}
\begin{center}
\begin{tabular}{|c|c|c|c|}
\hline
$\bm{m}$ & \textbf{Dataset Type} & \textbf{Model} & \textbf{EV Counts} \\
\hline
5 & Training & $TF_{5}$ & $E_{5} \in \{20,25,30,...,100\}$  \\
\hline
10 & Training & $TF_{10}$ & $E_{10} \in \{20,30,40,...,100\}$  \\
\hline
15 & Training & $TF_{15}$ & $E_{15} \in \{20,35,50,...,100\}$  \\
\hline
20 & Training & $TF_{20}$ & $E_{20} \in \{20,40,60,...,100\}$  \\
\hline
$-$ & Testing & $-$ & $E_{t} \in [20,100]\;/\;E_{5}$  \\
\hline
\end{tabular}
\label{dataset}
\end{center}
\end{table}

\begin{table}[b]
\caption{Prediction and Solution Performance of each trained transformer model}
\begin{center}
\begin{tabular}{|c|c|c|c|c|c|}
\hline
\textbf{Transformer}&\multicolumn{5}{|c|}{\textbf{Metric (\%)}} \\
\cline{2-6} 
\textbf{Model} & \textbf{\textit{$Acc_0$}} & \textbf{\textit{$Acc_1$}} & \textbf{\textit{$\bar{r}$}}& \textbf{\textit{$\bar{l}$}} & \textbf{\textit{$feas$}} \\
\hline
$TF_{5}$ & 99.84 & 63.02 & 89.00 & $-2.4 \times 10^{-4}$ & 92.00 \\
\hline
$TF_{10}$ & 99.84 & 63.60 & 92.65 & $6.6 \times 10^{-4}$ & 94.50 \\
\hline
$TF_{15}$ & 99.84 & 64.09 & 98.13 & $6.5 \times 10^{-4}$ & 100.00 \\
\hline
$TF_{20}$ & 99.84 & 63.73 & 98.18 & $1.1 \times 10^{-3}$ & 100.00 \\
\hline
\end{tabular}
\label{result}
\end{center}
\end{table}

Each transformer model is then evaluated using the metrics outlined in Section \ref{method} and their performance at assisting Gurobi to solve the EVJRS problem is
reported in Table \ref{result}. The results indicate that all models were able to learn the binary solutions relatively well, as reflected by the relatively high $ACC_0$ and $ ACC_1$. A notable observation is that the models trained on datasets with coarser granularity ($TF_{15}$ and $TF_{20}$) exhibit higher $ACC_1$ compared to those trained on datasets with finer granularity  ($TF_5$ and $TF_{10}$). The accuracies suggest that there is a trade-off between data granularity and data complexity. While finer granularity data (i.e., $TF_5$) provides the model access to more scenarios that can potentially allow the model to learn more precise case-by-case mapping, it can also increase training complexity, thereby hindering effective model convergence. In contrast, models trained on coarser granularity data (i.e., $TF_{20}$) can simplify the training process at the expense of learning general solutions that may be more suboptimal. This is further reflected in each models' $\bar{r}$, $\bar{l}$ and $feas$, where although $TF_5$ achieved the lowest objective values, it has the worst feasibility rate (consequently the worst computational time reduction). On the other hand, $TF_{10}$, $TF_{15}$ and $TF_{20}$ exhibit higher objective values but with higher feasibility rate, with $TF_{15}$ having a balance between high feasibility rate and low objective values. Thus, $TF_{15}$ will be the best option to assist Gurobi to solve the EVJRS problem.

\subsection{Benchmarking}

\begin{table}[t]
\caption{Prediction and Solution Performance of each trained CNN model}
\begin{center}
\begin{tabular}{|c|c|c|c|c|c|}
\hline
\textbf{CNN}&\multicolumn{5}{|c|}{\textbf{Metric (\%)}} \\
\cline{2-6} 
\textbf{Model} & \textbf{\textit{$Acc_0$}} & \textbf{\textit{$Acc_1$}} & \textbf{\textit{$\bar{r}$}}& \textbf{\textit{$\bar{l}$}} & \textbf{\textit{$feas$}} \\
\hline
$CNN_{5}$ & 99.88 & 59.62 & 97.83 & -0.01 & 99.50 \\
\hline
$CNN_{10}$ & 99.87 & 60.40 & 95.81 & 0.06 & 97.50 \\
\hline
$CNN_{15}$ & 99.87 & 61.12 & 91.48 & 0.29 & 95.50 \\
\hline
$CNN_{20}$ & 99.85 & 60.44 & 75.59 & 0.52 & 77.00 \\
\hline
\end{tabular}
\label{result1}
\end{center}
\end{table}

In this section, we compare the proposed transformer model with a convolutional neural network (CNN) model that incorporates a padding mechanism that is introduced in our previous work \cite{yap2025joint}. The padding mechanism enables any conventional DL models to handle problem instances with variable EV fleet sizes. The mechanism works by first estimating a maximum number of EVs that is expected for the EVJRS problem. A standard DL model, in this case CNN, is then designed based on the estimated upper limit. When met with instances with lesser EVs, zero padding can then be applied to the input and outputs corresponding to the missing EVs to ensure the dimensions do not change.

The performance of the CNN with padding is shown in Table \ref{result1}. When comparing the best performing transformer ($TF_{15}$) and CNN ($CNN_5$) model, the transformer is shown to be more advantageous as it has achieved 100\% feasibility, albeit with slight reduction in solution quality. In practical scenarios, an algorithm that consistently yields sub-optimal solutions is always more desirable than one that occasionally gives optimal feasible solutions. Furthermore, the proposed transformer offers additional advantage in scalability, as by design it is able to make predictions for problems with fleet sizes that were never encountered during training.

\section{Conclusion and Future Work} \label{conclude}

In this work, a fleet-size-agnostic transformer-based DL model has been developed to assist a commercial solver, Gurobi to solve a MIP-based, day-ahead EVJRS charging coordination problem. The EVJRS problem seeks to simultaneously optimise the routing and charging-discharging decisions of an EV fleet operating within a DN and TN. The proposed framework improves the tractability of the problem by leveraging DL's fast inference time and trained a transformer model to predict the problem's optimal binary solutions, thereby effectively pruning the solution search space for the solver. The transformer model is designed to handle problems with EV fleet of any sizes even if it was never seen during training, thus reducing the number of re-training required and improving the practicality of the approach. Simulations were performed on the IEEE 33-bus system coupled with the Nguyen-Dupuis TN with fleet size of 20 to 100 EVs to evaluate the model's performance. When compared to Gurobi, the proposed approach reduced solution runtime by 98.1\% on average, while retaining 100.0\% feasibility and losing less than 0.0007\% solution quality, all without re-training the model. In the future, it would be valuable to investigate the how well the model generalises to problems with large EV fleets while only trained on small fleets, since solving large instances to obtain the ground truth label can be very computationally expensive. 

\bibliography{references}{}
\bibliographystyle{IEEEtran}

\end{document}